\documentclass[11pt,a4paper]{article}
\usepackage[hyperref]{acl2021}
\usepackage{times}
\usepackage{latexsym}

\usepackage{microtype}

\aclfinalcopy % Uncomment this line for the final submission
\usepackage{soul}
\usepackage{url}
\usepackage{graphicx}
\usepackage{amsmath}
\usepackage{amsthm}
\usepackage{booktabs}
\usepackage{algorithm}
\usepackage{algorithmic}
\usepackage{float}
\usepackage{subfigure}
\usepackage{xcolor}
\usepackage{mathtools}
\usepackage{amsmath,bm}
\usepackage{makecell}
\usepackage{multirow}
\usepackage{multicol}
\usepackage{array}
\usepackage{comment}
\usepackage{amsfonts,amssymb}
\usepackage{textcomp}
\usepackage{fmtcount}
\usepackage{flushend}
\usepackage{thmtools}
\usepackage{xspace}

\usepackage{balance}

\usepackage{CJKutf8}

\newcommand{\eg}{\hbox{\emph{e.g.}}\xspace}

\newcommand{\ie}{\hbox{\emph{i.e.}}\xspace}
\newcommand{\wrt}{\hbox{\emph{w.r.t.}}\xspace}

\newcommand{\tabincell}[2]{\begin{tabular}{@{}#1@{}}#2\end{tabular}}

\title{OntoED: Low-resource Event Detection with Ontology Embedding}

\author{
	Shumin Deng$^{1,2 *}$, Ningyu Zhang$^{1,2}$\thanks{$\quad$ Equal Contribution.}~, Luoqiu Li$^{1,2}$, 
	Hui Chen$^{3}$, Huaixiao Tou$^{3}$, \\
	\textbf{
		Mosha Chen$^{3}$, Fei Huang$^{3}$,
		Huajun Chen$^{1,2}$\thanks{$\quad$ Corresponding Author.} 
	} \vspace{1.0mm} \\
	$^1$Zhejiang University \& AZFT Joint Lab for Knowledge Engine, China \\
	$^2$Hangzhou Innovation Center, Zhejiang University, China \\
	$^3$Alibaba Group, China \\
	\fontsize{11}{10}\selectfont \{231sm, zhangningyu, 3160102409\}@zju.edu.cn, \{weidu.ch, huaixiao.thx\}@alibaba-inc.com, \\
	\fontsize{11}{10}\selectfont \{chenmosha.cms, f.huang\}@alibaba-inc.com, huajunsir@zju.edu.cn \\
}
\date{}

\begin{document}

\maketitle

\begin{abstract}
Event Detection (ED) aims to identify event trigger words from a given text and classify it into an event type. 
Most of current methods to ED rely heavily on training instances, and almost ignore the correlation of event types. Hence, they tend to suffer from data scarcity and fail to handle new unseen event types. 
To address these problems, we formulate ED as a process of event ontology population: linking event instances to pre-defined event types in event ontology, and propose a novel ED framework entitled OntoED with ontology embedding. 
We enrich event ontology with linkages among event types, and further induce more event-event correlations. 
Based on the event ontology, OntoED can leverage and propagate correlation knowledge, particularly from data-rich to data-poor event types. 
Furthermore, OntoED can be applied to new unseen event types, by establishing linkages to existing ones.  
Experiments indicate that OntoED is more predominant and robust than previous approaches to ED, especially in data-scarce scenarios.  
% outperforms previous methods of ED, particularly with low-resource training data regimes. 
% without additional manual annotations
% , and they can establish linkages to existing one in the event ontology.

\end{abstract}
% a new learning paradigm of ED

\iffalse
\bibliographystyle{acl_natbib}
\bibliography{acl2021}
\fi
% !TEX root = ./acl2021.tex

\section{Introduction}
\label{sec:intro}
% ------Intro of ED tasks-----
Event Detection (ED) \cite{ACL2015_DMCNN} is the task to extract structure information of events from unstructured texts. 
For example, in the event mention ‘‘\emph{Jack is married to the Iraqi microbiologist known as Dr. Germ.}'', an ED model should identify the event type as ‘\emph{Marry}' where the word ‘\emph{married}' triggers the event. 
The extracted events with canonical structure facilitate various social applications, such as biomedical science \cite{NAACL2019_BiomedicalEE,EMNLP2020_BiomedicalEE}, financial analysis \cite{WWW2019KGTA_Event4TSPA,IJCAI2020_FinancialED}, fake news detection \cite{KDD2018_ED4News,AIAI2020_ED4News} and so on.

\begin{figure}[!htbp]
  \centering
  \includegraphics[width=0.92\linewidth]{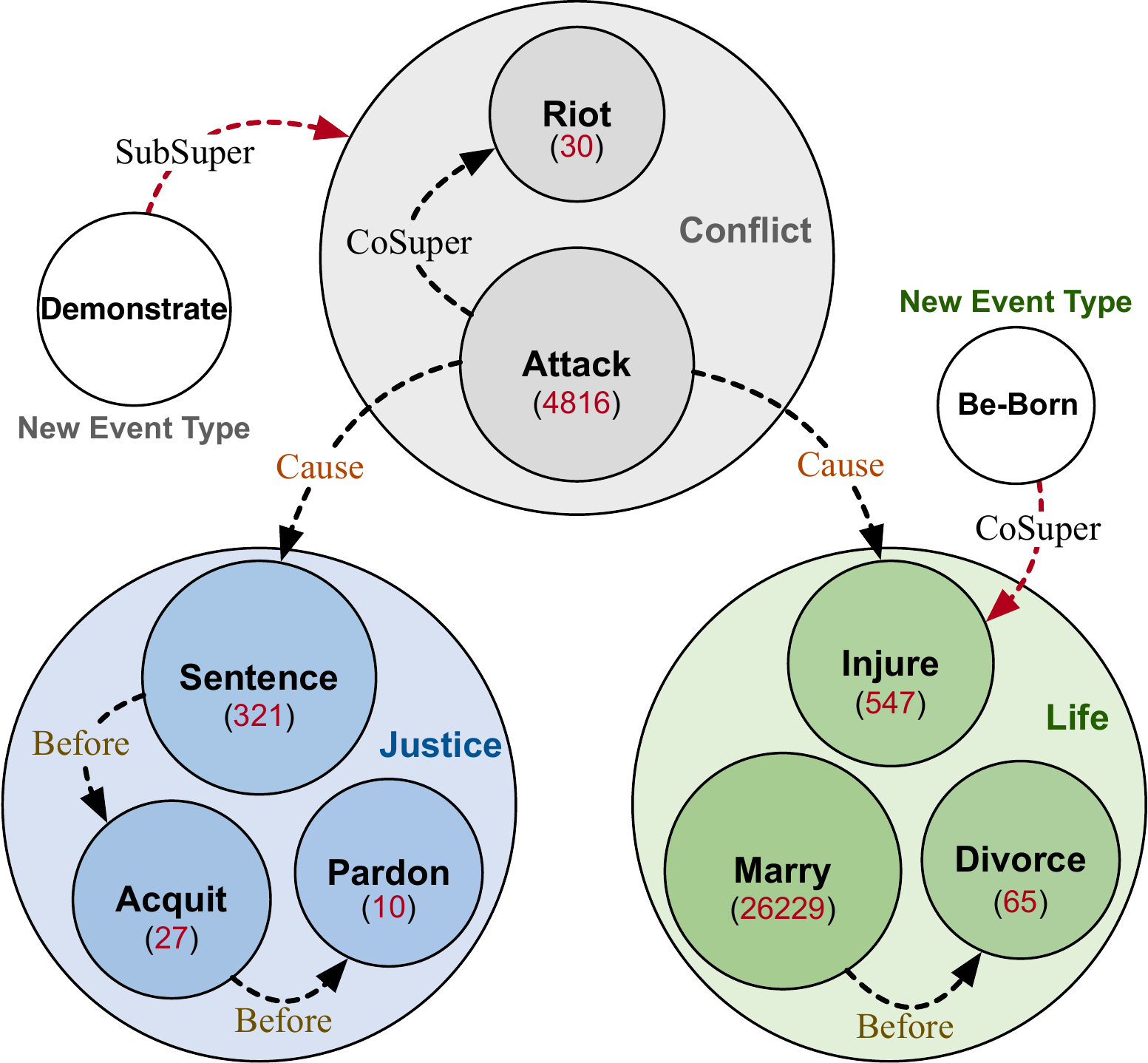}
  \caption{Low-resource Event Detection \wrt Event Correlation in \emph{FewEvent} \cite{WSDM2020_MetaL-EE_DMBPN} Dataset. 
  \label{fig:intro_motivation} }
  \vspace{-3mm}
\end{figure}

% ------low-resource challenges in ED-----
As a non-trivial task, ED suffers from the low-resource issues. 
On the one hand, the maldistribution of samples is quite serious in ED benchmark datasets, \eg, \emph{FewEvent} \cite{WSDM2020_MetaL-EE_DMBPN} and \emph{MAVEN} \cite{EMNLP2020_MAVEN}, where a large portion of event types contain relatively few training instances. As shown in Figure~\ref{fig:intro_motivation}, the sample size of two event types \emph{Attack} and \emph{Riot} differs greatly (4816 \& 30). 
In low-resource scenarios, supervised ED models \cite{ACL2015_DMCNN,NAACL2016_JRNN,EMNLP2018_JMEE} are prone to overfitting since they require sufficient training instances for all event types. 
On the other hand, real-world applications tend to be open and evolve promptly, and accordingly there can be numerous new unseen event types. 
Handling new event types may even entail starting over, without being able to re-use annotations from previous ones \cite{ACL2018_ZSEE}. 

% ------proposal of event ontology-----
Regarding low-resource ED, \citet{ACL2018_ZSEE} take a fresh look at ED, by mapping each event mention to a specific type in a target event ontology, which can train from few seen event types and then transfer knowledge to new unseen ones. However, the event ontology here merely considers the intra-structure for each event mention and event type. 
In this paper, we enrich the event ontology with more inter-structures of event types, such as temporal, causal and hierarchical event-event relations \cite{ACL2018_MATRES,EMNLP2020_ERE_H-T}. 
For example, as seen in Figure~\ref{fig:intro_motivation}, $\emph{Attack} \xrightarrow{ \textsc{Cause} } \emph{Sentence}$, $\emph{Sentence} \xrightarrow{ \textsc{Before} } \emph{Acquit}$, $\emph{Attack} \xrightarrow{ \textsc{CoSuper} } \emph{Riot}$. 
Our key intention is to fully utilize the event ontology and leverage correlation knowledge from data-rich event types (\ie, \emph{Attack}) to data-poor ones (\ie, \emph{Sentence}, \emph{Acquit} and \emph{Riot}). 
Besides, new event types (\ie, \emph{Be-Born}) can be learned with correlation (\ie, $\textsc{CoSuper}$) of existing ones (\ie, \emph{Injure}). 

As the first attempt to construct such event ontology, we propose a novel ED framework with ontology embedding called OntoED. 
First, we establish the initial event ontology with event instances and types. We capture semantic features and relations of event instances with BERT \cite{NAACL2019_BERT} and utilize prototypes \cite{NIPS2017_PN} to represent event types, where a prototype is the average of its instance embeddings. 
Second, we extend the event ontology with event-event relations based on extracted relations among event instances, and then learn ontology embedding by aggregating neighbor prototypes for each prototype \wrt correlations among event types. In this way, semantically similar event types in vector space will be closer, thus, improving the discrimination of dissimilar event types. 
Third, we design an event correlation inference mechanism to induce new event correlations based on symbolic rules, \eg, (\emph{Sentence}, $\textsc{Before}$, \emph{Acquit}) $\land$ (\emph{Acquit}, $\textsc{Before}$, \emph{Pardon}) $\rightarrow$ (\emph{Sentence}, $\textsc{Before}$, \emph{Pardon}). Thus, we can induce new event-event relations to further enrich the event ontology. 
To the best of our knowledge, it is the first work to explicitly model correlations among event types with event ontology in low-resource ED. 

Our contributions can be summarized as follows: 
\begin{itemize}

    \item  We study the low-resource event detection problem and propose a novel ontology-based model, OntoED, that encodes intra and inter structures of events. 

    \item We provide a novel ED framework based on ontology embedding with event correlations, which interoperates symbolic rules with popular deep neural networks. 

    \item We build a new dataset \texttt{OntoEvent} for ED. Extensive experimental results demonstrate that our model can achieve better performance on the overall, few-shot, and zero-shot setting.

\end{itemize}

\iffalse
\bibliographystyle{acl_natbib}
\bibliography{acl2021}
\fi
% !TEX root = ./acl2021.tex

\section{Related Work}
\label{sec:related_work}

% \textbf{Low-resource Event Detection}. 
Traditional approaches to ED are mostly based on neural networks \cite{ACL2015_DMCNN,NAACL2016_JRNN,EMNLP2018_JMEE,EMNLP2019_AD-DMBERT,EMNLP2019_MOGANED,EMNLP2020_EE-GCN,COLING2020_PAJHEE,ACL2021_MLBiNet}, and ignore correlation knowledge of event types, especially in low-resource scenarios. 
Most previous low-resource ED methods \cite{EMNLP2016_DS-EE} have been based on \emph{supervised learning}. 
However, supervised-based methods are too dependent on data, and fail to be applied to new types without additional annotation efforts. 
Another popular methods for low-resource ED are based on \emph{meta learning}. 
\citet{WSDM2020_MetaL-EE_DMBPN,ACL2020-NUSE_MatchEE,ACL2021_KEFSED} reformulate ED as a few-shot learning problem to extend ED with limited labeled samples to new event types, and propose to resolve few-shot ED with meta learning. 
Besides, \emph{knowledge enhancement} and \emph{transfer learning} are applied to tackle low-resource ED problems. 
\citet{ACL2020_EKD} leverage open-domain trigger knowledge to address long-tail issues in ED. 
\citet{EMNLP2020_RCEE,EMNLP2020_QAEE} propose to handle few-shot and zero-shot ED tasks by casting it as a machine reading comprehension problem. 
\citet{ACL2018_ZSEE} propose to tackle zero-shot ED problem by mapping each event mention to a specific type in a target event ontology. Note that \citet{ACL2018_ZSEE} establish the event ontology merely with intra-structure of events, while we extend it with inter-structure of event correlations.  
Though these methods are suitable for low-resource scenarios, they mostly ignore implicit correlation among event types and lack reasoning ability. 

% \noindent \textbf{Ontology-based Information Extraction}. 
In order to utilize correlation knowledge among event types, \citet{EMNLP2020_PathLM} propose a new event graph schema, where two event types are connected through multiple paths involving entities. However, it requires various annotations of entities and entity-entity relations, which is complicated and demanding. 
Different from \citet{EMNLP2020_PathLM}, we propose to revisit the ED task as an ontology learning process, inspired by relation extraction (RE) tasks based on ontology and logic-based learning. 
\citet{J2018_OntoILPER,J2019_OntoILPER} present a logic-based relational learning approach to RE that uses inductive logic programming for generating information extraction (IE) models in the form of symbolic rules, demonstrating that ontology-based IE approaches are advantageous in capturing correlation among classes, and succeed in symbolic reasoning.

\iffalse
\bibliographystyle{acl_natbib}
\bibliography{acl2021}
\fi
% !TEX root = ./acl2021.tex

\section{Methodology}
\label{sec:method}

\subsection{Problem Formulation}

We revisit the event detection task as an iterative process of event ontology population. 
Given an event ontology $\mathcal{O}$ with an event type set $\mathcal{E} = \{ e_i | i \in [1, N_e] \}$, and corpus $\mathcal{T} = \{ X_i | i \in [1, K] \}$ that contains $K$ instances, the goal of event ontology population is to establish proper linkages between event types and instances. Specifically, each instance $X_i$ in $\mathcal{T}$ is denoted as a token sequence $X_i = \{x_i^j | j \in [1, L]\}$ with maximum $L$ tokens, where the event trigger $x_i^t$ are annotated. We expect to predict the index $t$ ($1 \leq t \leq L$) and the event label $e_i$ for each instance respectively. 
% of the trigger candidate

Besides, we utilize a multi-faceted event-event relation set $\mathcal{R} = \mathcal{R}_H \sqcup \mathcal{R}_T \sqcup \mathcal{R}_C$ for event ontology population and learning. Thereinto, 
$\mathcal{R}_H = \{ \textsc{SubSuper}, \textsc{SuperSub}, \textsc{CoSuper}\footnote{($e_i$, \textsc{CoSuper}, $e_j$): $e_i$ and $e_j$ has the same super type.} \}$ denotes a set of relation labels defined in the subevent relation extraction task \cite{EMNLP2020_ERE_H-T,EMNLP2020_ERE_H}.
$\mathcal{R}_T = \{ \textsc{Before}, \textsc{After}, \textsc{Equal}\footnote{($e_i$, \textsc{Equal}, $e_j$): $e_i$ and $e_j$ happens simultaneously.} \}$ denotes a set of temporal relations \cite{EMNLP2020_ERE_T}.
$\mathcal{R}_C = \{ \textsc{Cause}, \textsc{CausedBy} \}$ denotes a set of causal relations \cite{ACL2018_MATRES}. 
% ACL2014_TimeBank-Dense,

\subsection{Model Overview}
\begin{figure}[!htbp]
  \centering
  \includegraphics[width=0.98\linewidth]{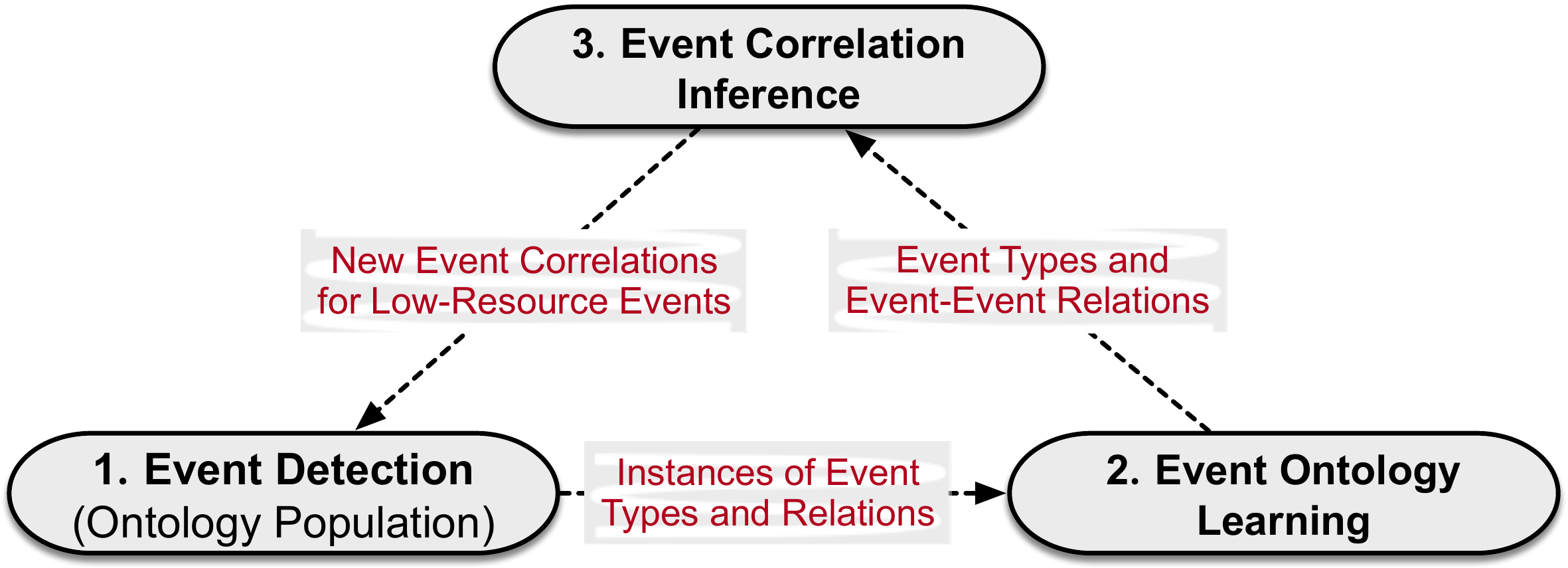}
  \caption{Overview of our proposed OntoED. 
  \label{fig:model_overview} }
  % \vspace{-4mm}
\end{figure}

In this paper, we propose a general framework called \emph{OntoED} with three modules: (1) Event Detection (Ontology Population), (2) Event Ontology Learning, and (3) Event Correlation Inference. Figure~\ref{fig:model_overview} shows the key idea of the three modules.
% which extracts events and correlations for low-resource ED. 

% \noindent 
\emph{Event Detection} aims at identifying the event trigger $x_i^t$ and type $e_i$ for each input tokens $X_i$, and then identify relations among event instances. The average instance embedding of each type is calculated as the primitive event prototype. 

\emph{Event Ontology Learning} aims to obtain event ontology embedding with the correlation of event prototypes, based on the relations among event types derived from instances. 

\emph{Event Correlation Inference} seeks to infer new event correlations based on existing event-event relations, so as to obtain a solid event ontology. 

\begin{figure*}[!htbp]
  \centering
  \includegraphics[width=0.98\linewidth]{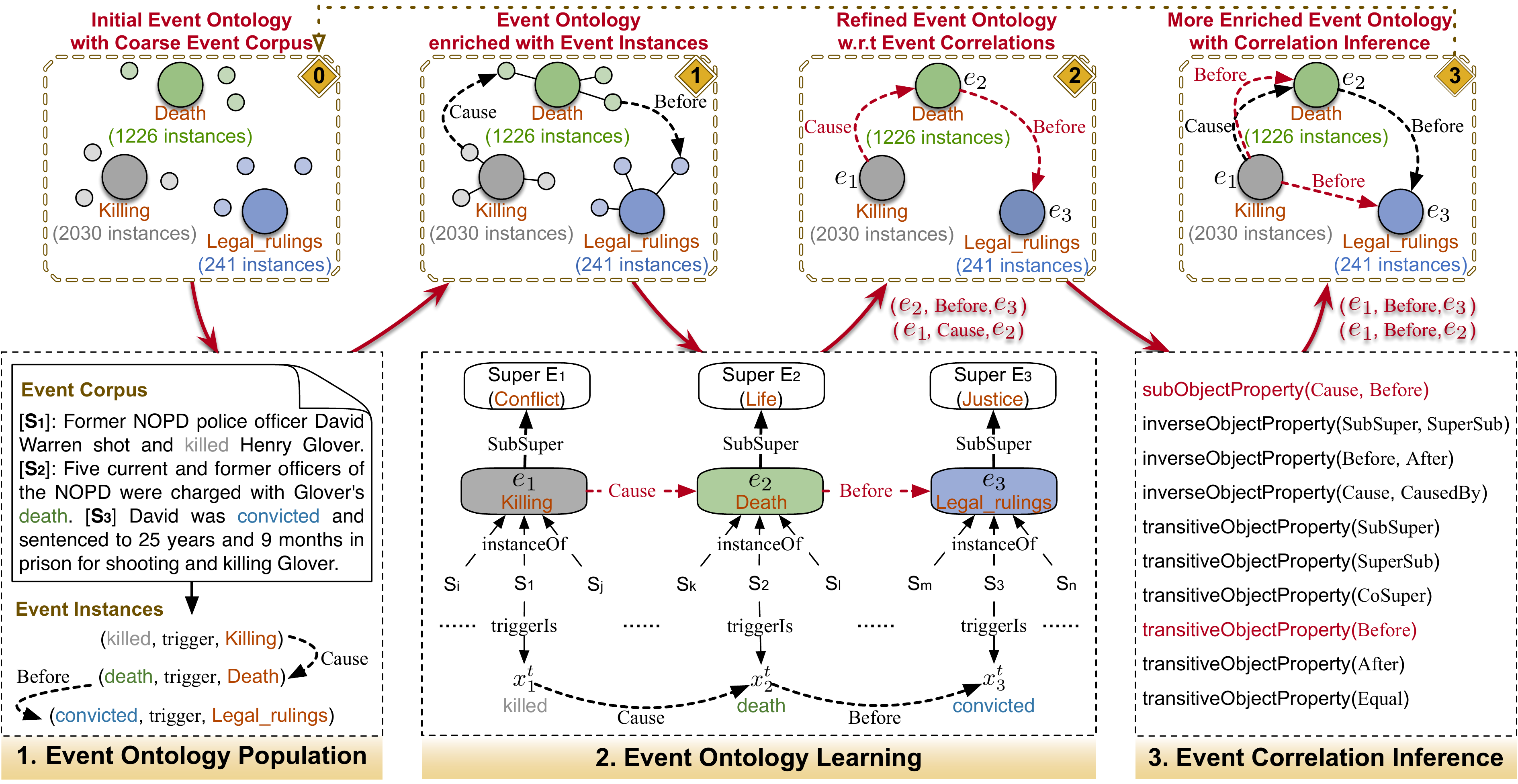}
  \caption{Detailed example for the process of OntoED. 
  Note that we ignore instance nodes in No.2 and No.3 event ontology for space limit. 
  \textbf{Step 1}: \textbf{Event Detection (Ontology Population)} connect event types with instances, given the initial event ontology with coarse corpus.
  \textbf{Step 2}: \textbf{Event Ontology Learning} establish correlations among event types, given the event ontology enriched with instances. 
  \textbf{Step 3}: \textbf{Event Correlation Inference} induce more event correlations based on existing event-event relations, \eg, $(e_1, \textsc{Cause}, e_2) \to (e_1, \textsc{Before}, e_2)$, and $(e_1, \textsc{Before}, e_2) \land (e_2, \textsc{Before}, e_3) \to (e_1, \textsc{Before}, e_3)$. 
  \label{fig:model_overview_detail} }
  % \vspace{-3mm}
\end{figure*}

The detailed architecture of OntoED with running examples is illustrated in Figure~\ref{fig:model_overview_detail}.

\subsection{Event Detection (Ontology Population)}
The input of ED is an initial event ontology with event types $\mathcal{E}$ and coarse corpus $\mathcal{T}$. 

\textbf{Instance Encoder}. 
Given a token sequence $X_i = \{ x_i^1, \cdots, x_i^L \}$ with trigger $x_i^t$, we use a pre-trained BERT \cite{NAACL2019_BERT} to get a contextual representation $\bm{X}_i^t$ for $x_i^t$, and use the token embedding of [CLS] as the contextual representation $\bm{X}_i$ for $X_i$. Note that the instance encoder is pluggable, and can be replaced as other models followed by \cite{WSDM2020_MetaL-EE_DMBPN,EMNLP2020_EE-GCN}. 

\textbf{Class Encoder}. 
We then represent event types as \emph{prototypes} \cite{NIPS2017_PN}, as it is proven to be robust for low-resource ED \cite{WSDM2020_MetaL-EE_DMBPN}. 
% We select this method as it is proven to be robust in low-resource event detection scenarios \cite{WSDM2020_MetaL-EE_DMBPN}. 

Initially, event types have no correlation with others, thus we require to compute the \emph{prototype} $\bm{P}_k$ for $e_k \in \mathcal{E}$ by averaging its instance embeddings: 
\begin{equation}
% \small
  \bm{P}_k = \frac{1}{N_k} \sum\nolimits_{i=1}^{N_k} \bm{X}_i
\label{eq:event_proto_ed}
\end{equation}
where $N_k$ is the instance number of $e_k$. 
Afterward, event prototypes will be induced from the module of event correlation inference, as shown in Figure~\ref{fig:model_overview_detail}. 

\textbf{Event Detector}. 
Given embeddings of a token sequence, we treat each token as an event trigger candidate and then compute probability of the corresponding event type for event trigger candidate $x_i^t$, denoted by: 
\begin{equation}
% \small
  P(y = e_k) = \frac{exp(- \| \bm{X}_i^t - \bm{P}_k \| )}{\sum\nolimits_{j=1}^{N_e} exp(- \| \bm{X}_i^t - \bm{P}_j \| )}
\label{eq:prob_ed}
\end{equation}
where $\| \cdot \|$ denotes Euclidean distance, and $N_e = |\mathcal{E}|$ denotes the number of event types.

As general, we adopt cross entropy as the loss function for event detection, denoted by: 
\begin{equation}
% \small
  \mathcal{L}_{ED} = - \sum\nolimits_{k=1}^{N_e} y \log P(y = e_k)
\label{eq:loss_ed}
\end{equation}
where $y$ is the ground-truth label for $x_i^t$. 

\textbf{Instance Relation Extractor}. 
For each event instance pair $(X_i, X_j)$, we adopt a comprehensive way to model embedding interactions \cite{ACL2020_EventPairRep}, denoted by $\bm{X}_{ij}^p = [ \bm{X}_i, \bm{X}_j, \bm{X}_i \odot \bm{X}_j, \bm{X}_i - \bm{X}_j ]$,  
 % in a document with a set of token sequences $\{ X_1, \cdots, X_{N_d} \}$
% denoted by: 
% % the concatenation of $\bm{X}_i$ and $\bm{X}_j$, along with their element-wise Hadamard product and subtraction.
% \begin{equation}
% % \small
%   \bm{X}_{ij}^p = [ \bm{X}_i, \bm{X}_j, \bm{X}_i \odot \bm{X}_j, \bm{X}_i - \bm{X}_j ]
% \label{eq:event_pair_rep}
% \end{equation}
where $[ \cdot, \cdot ]$ denotes a vector concatenation, and $\odot$ is the element-wise Hadamard product. 

We then calculate the probability $P(y = r_k)$ of relation $r_k \in \mathcal{R}$ between $(X_i, X_j)$ by softmax. 
Generally, we adopt cross entropy as the loss function for instance relation extraction, denoted by: 
\begin{equation}
% \small
  \mathcal{L}_{RE} = - \sum\nolimits_{k=1}^{N_r} y \log P(y = r_k)
\label{eq:loss_re}
\end{equation}
where $y$ is the ground-truth for $(X_i, X_j)$, and $N_r = |\mathcal{R}|$ denotes the number of event-event relations. 

Overall, the loss function for event detection (ontology population) is calculated by: 
\begin{equation}
% \small
  \mathcal{L}_{OP} = \gamma \mathcal{L}_{ED} + (1 - \gamma) \mathcal{L}_{RE}
\label{eq:loss_op}
\end{equation}
where $\gamma$ is a hyperparameter.

\begin{table*}[!htbp]
\centering
\small
%\footnotesize
%\scriptsize
%\tiny
\resizebox{\linewidth}{!}{
\begin{tabular}{ c | c | c | c }
\toprule

\textbf{Object Property Axioms} 
	& \textbf{Rule Form}
	& \textbf{According to Linear Map Assumption} 
	& \textbf{Relation Constraint} \\

\midrule

% $\mathtt{SymmetricOP}(r)$ 
% 	& $(e_j, r, e_i) \gets (e_i, r, e_j)$
% 	& $\bm{P}_j \bm{M}_r = \bm{P}_i$; $\bm{P}_i\bm{M}_r=\bm{P}_j $
% 	& $ \bm{M}_r\bm{M}_r = \bm{I}$ \\

% \hline

$\mathtt{subOP}(r_1, r_2)$ 
	& $(e_i, r_2, e_j) \gets (e_i, r_1, e_j)$
	& $\bm{P}_i \bm{M}_{r_2} = \bm{P}_j,\; \bm{P}_i \bm{M}_{r_1} = \bm{P}_j$
	& $\bm{M}_{r_1} = \bm{M}_{r_2}$ \\

% \hline

$\mathtt{inverseOP}(r_1, r_2)$
	& $(e_i, r_1, e_j) \gets (e_j, r_2, e_i)$
	& $\bm{P}_i \bm{M}_{r_1} = \bm{P}_j,\; \bm{P}_j \bm{M}_{r_2} = \bm{P}_i$ 
	& $\bm{M}_{r_1} \bm{M}_{r_2} = \bm{I}$ \\

% \hline

$\mathtt{transitiveOP}(r)$ 
	& $(e_i, r, e_k) \gets (e_i, r, e_j), (e_j, r, e_k)$
	& $\bm{P}_i \bm{M}_r = \bm{P}_k,\; \bm{P}_i \bm{M}_r = \bm{P}_j,\; \bm{P}_j \bm{M}_r = \bm{P}_k$
	& $\bm{M}_r \bm{M}_r = \bm{M}_r$ \\

% \hline

% $\mathtt{subOP}(r_1, r_2)$
% 	& $(e_i, r_2, e_j) \gets (e_i, r_1, e_j)$  
% 	& $\bm{P}_i \bm{M}_{r_2} = \bm{P}_j,\; \bm{P}_i \bm{M}_{r_1} = \bm{P}_j$
% 	& $\bm{M}_{r_1} = \bm{M}_{r_2}$  \\ 
% \hline

\bottomrule
\end{tabular}
}

\caption{ 
\small{
	Three types of object property expression axioms. $\mathtt{OP}$ is the short for $\mathtt{ObjectProperty}$. 
	%  selected from OWL2
	% $\bm{P}$ and $\bm{M}_r$ denote entity and relation embeddings respectively.
	$\bm{I}$ is an identity matrix. 
}
}
\label{tab:object_properties}

\end{table*}

\subsection{Event Ontology Learning}
% As mentioned before, we represent event types as prototypes with type encoder. However, these event prototypes are constructed from instances. We propose to formalize them into an event ontology, for ease of learning more proper prototypes. 

\textbf{Ontology Completion}. 
We complete event ontology $\mathcal{O}$ with both intra and inter structure of events. We normatively link event instances $\mathcal{T}$ to event types $\mathcal{E}$, and establish correlations among event types based on linkages among event instances. 

\emph{Instance-to-class Linking}. 
Given a sentence $S_i$ (formalized as a token sequence $X_i$) with a trigger $x_i^t$ of an event instance, we link these information to its corresponding event type $e_i$ with normative triples: ($S_i$, triggerIs, $x_i^t$) and ($S_i$, instanceOf, $e_i$). 

\emph{Class-to-class Linking}. 
Given an event instance pair $(X_i, X_j)$ with a relation $r$, we upgrade the instance correlation to corresponding event types, denoted by $(e_i, r, e_j)$.  
Besides, we link each event subtype to its corresponding supertype\footnote{The supertypes and its corresponding subtypes in this paper are pre-defined and will be introduced in appendix.} with a $\textsc{SubSuper}$ relation ($\textsc{SuperSub}$ in reverse), and we link each event subtype pair having the same supertype with a $\textsc{CoSuper}$ relation.

\textbf{Ontology Embedding}. 
We represent the event ontology considering both instances and correlations for each event type. 
Specifically, given a triple $\ell = (e_h, r, e_t) \in \mathcal{O}$, we propagate the prototype $\bm{P}_h$ of head event type $e_h$ to prototype $\bm{P}_t$ of tail event type $e_t$ with a relation transformation matrix $\bm{M}_r \in \mathbb{R}^{d \times d}$. We select a matrix to embed $r$ as it shows great robustness to model relations in low-resource senarios \cite{WWW2019_IterE}. 
We then aggregate propagation from all head event types by 
\begin{equation}
% \small
  \bm{P}_t^{\ast} = \sum\limits_{ (e_h^i, r_i, e_t) \in \mathcal{O}_{\ell}} \bm{P}_h^i \bm{M}_{r_i}
\label{eq:proto_prop_agg}
\end{equation}
where $\mathcal{O}_{\ell}$ is all one-hop neighbor triples of $e_t$ in $\mathcal{O}$.
% where $\mathcal{O}_{\ell}$ denotes all the neighbor triples of $e_t$ with one hop in $\mathcal{O}$. 
% % $\bm{M}_{r_i} \in \mathbb{R}^{d \times d}$ is a relation transformation matrix for $r_i \in \mathcal{R}$. 

The prototype $\bm{P}_t$ of $e_t$ in $\ell$ after propagation is a weighted average of $\bm{P}_t$ and $\bm{P}_t^{\ast}$ with weight $\lambda \in [0, 1]$, denoted by: 
\begin{equation}
% \small
  \bm{P}_t = \lambda \bm{P}_t + (1 - \lambda) \bm{P}_t^{\ast}
\label{eq:proto_prop_final}
\end{equation}

We calculate the possibility that $r$ is the relation between $e_h$ and $e_t$ with a truth value for $(e_h, r, e_t)$: 
$\phi(e_h, r, e_t) = sim ( \bm{P}_h \bm{M}_r, \bm{P}_t ) = \sigma ( \bm{P}_h^{\top} \bm{M}_r \bm{P}_t )$, where 
% \begin{equation}
% % \small
%   \phi(e_h, r, e_t) = sim ( \bm{P}_h \bm{M}_r, \bm{P}_t ) = \sigma ( \bm{P}_h^{\top} \bm{M}_r \bm{P}_t )
% \label{eq:proto_prop_triple_value}
% \end{equation}
$\sigma$ is sigmoid function, and the similarity between $\bm{P}_h \bm{M}_r$ and $\bm{P}_t$ is evaluated via dot product. 

Overall, the loss fuction for event ontology learning is defined by: 
\begin{equation}
% \small
  \mathcal{L}_{OL} = - \sum\limits_{(e_h, r, e_t) \in \mathcal{O}} y \log \phi(e_h, r, e_t)
\label{eq:loss_event_onto_final}
\end{equation}
and $y$ denotes the ground-truth label for $(e_h, r, e_t)$.

\subsection{Event Correlation Inference}
Given the event ontology with correlations among event types, we infer new event correlations based on existing ones. 
% , in order to obtain more enriched event ontology
% and the grounding $g$ considered in this paper
To be specific, we utilize the grounding $g$ to infer new event correlation triples, which can be generalized as the following form: 
\begin{equation}
% \small
  (e_h^I, r^I, e_t^I) \gets (e_h^1, r^1, e_t^1), \cdots, (e_h^n, r^n, e_t^n)
\label{eq:event_rel_grounding}
\end{equation}
where the right side event triples $(e_h^k, r^k, e_t^k) \in \mathcal{O}$ with $k \in [1, n]$ have already existed in $\mathcal{O}$ and $(e_h^I, r^I, e_t^I) \notin \mathcal{O}$ is new inferred triples to be added. 

To compute the truth value of the grounding $g$, we select three object properties ($\mathtt{OP}$) of relations defined in OWL2\footnote{\url{https://www.w3.org/TR/owl2-profiles/}} Web Ontology Language: $\mathtt{subOP}$, $\mathtt{inverseOP}$, and $\mathtt{transitiveOP}$, and then learn matrics of relations from linear map assumption \cite{WWW2019_IterE}, presented in Table~\ref{tab:object_properties}. 
\citet{EMNLP2020_ERE_H-T,ACL2018_MATRES} have defined some conjunctive constraints of relations between the event pair, we translate them into object property axioms, shown in Table~\ref{tab:object_properties_for_relations}. 

\begin{table}[!htbp]
\centering
\small
%\footnotesize
%\scriptsize
%\tiny
\resizebox{\linewidth}{!}{
\begin{tabular}{ c | c }
\toprule

\textbf{Object Property Axioms} 
	& \textbf{Instances of Relation / Relation Pair} \\

\midrule

	$\mathtt{subOP}(r_1, r_2)$ & ($\textsc{Cause}$, $\textsc{Before}$) \\

\midrule

	$\mathtt{inverseOP}(r_1, r_2)$ & 
	\tabincell{c}{ ($\textsc{SubSuper}$, $\textsc{SuperSub}$), \\ ($\textsc{Before}$, $\textsc{After}$), ($\textsc{Cause}$, $\textsc{Cause}$dBy) } \\

\midrule

	$\mathtt{transitiveOP}(r)$ & 
	\tabincell{c}{ $\textsc{SubSuper}$, $\textsc{SuperSub}$, $\textsc{CoSuper}$, \\ $\textsc{Before}$, $\textsc{After}$, $\textsc{Equal}$ } \\

\bottomrule
\end{tabular}
}

\caption{
	Groundings of three object properties in $\mathcal{O}$. 
}
\label{tab:object_properties_for_relations}

\end{table}

Assuming that $\bm{M}_r^{\dagger}$ and $\bm{M}_r^{\ddagger}$ denotes the relation set on left and right of Eq~\eqref{eq:event_rel_grounding} respectively, they are matrices either from a single matrix or a product of two matrices. As relation constraints are derived from ideal linear map assumption (the 3rd column in Table~\ref{tab:object_properties}), $\bm{M}_r^{\dagger}$ and $\bm{M}_r^{\ddagger}$ are usually unequal but similar during training. Thus, the normalized truth value $\mathcal{F}_p$ of $g$ can be calculated based on relation constraints (the 4th column in Table~\ref{tab:object_properties}): 
% \begin{equation}
% \small
$$
  \mathcal{F}_p^{'} = \| \bm{M}_r^{\dagger} - \bm{M}_r^{\ddagger} \|_F, ~ ~ 
  \mathcal{F}_p = \frac{ \mathcal{F}_p^{max} - \mathcal{F}_p^{'}}{ \mathcal{F}_p^{max} - \mathcal{F}_p^{min}  }
$$
% \label{eq:event_rel_infer_grounding_value}
% \end{equation}
where $\| \cdot \|_F$ denotes Frobenius norm, and subscript $p$ respectively denotes one of the three object properties. $\mathcal{F}_p^{max}$ and $\mathcal{F}_p^{min}$ is a the maximum and minimum Frobenius norm score. $\mathcal{F}_p \in [0, 1]$ is the truth value for the grounding $g$ and the higher $\mathcal{F}_p$ means the more confident that $g$ is valid.

The loss function for new event correlation inference is defined by:
\begin{equation}
% \small
\begin{aligned}
  \mathcal{L}_{ER} = & - \psi_S \sum\limits_{i \in \mathcal{G}^{(S)}} \log\mathcal{F}_p^i - \psi_V \sum\limits_{j \in \mathcal{G}^{(V)}} \log\mathcal{F}_p^j \\
  & - \psi_T \sum\limits_{k \in \mathcal{G}^{(T)}} \log\mathcal{F}_p^k
\label{eq:loss_er}
\end{aligned} 
\end{equation}
$\mathcal{G}^{(\cdot)}$ denotes all groundings \wrt $\mathtt{subOP}$ ($S$), $\mathtt{inverseOP}$ ($V$), and $\mathtt{transitiveOP}$ ($T$). $\psi_S$, $\psi_V$, and $\psi_T$ are hyperparameters for the loss of three object properties respectively. 

As a whole, the final loss function for OntoED is denoted by: 
\begin{equation}
% \small
  \mathcal{L} = 
  \alpha \mathcal{L}_{OP} + 
  \beta \mathcal{L}_{OL} + 
  \mathcal{L}_{ER}
\label{eq:loss_final}
\end{equation} 
where $\alpha$ and $\beta$ are hyperparameters for the loss of event ontology population (Eq \eqref{eq:loss_op}) and event ontology learning (Eq \eqref{eq:loss_event_onto_final}) respectively.

\iffalse
\bibliographystyle{acl_natbib}
\bibliography{acl2021}
\fi
% !TEX root = ./acl2021.tex

\section{Experiments}
\label{sec:experiment}
The experiments seek to: 
(1) demonstrate that OntoED with ontology embedding can benefit both standard and low-resource ED, and
(2) assess the effectiveness of different modules in OntoED and provide error analysis. 
To this end, we verify the effectiveness of OntoED in three types of evaluation: 
(1) \emph{Overall Evaluation}, 
(2) \emph{Few-shot Evaluation}, and 
(3) \emph{Zero-shot Evaluation}.

\subsection{Datasets}
As none of present datasets for ED is annotated with relations among events, 
we propose a new ED dataset namely \texttt{OntoEvent} with event correlations. 
It contains 13 supertypes with 100 subtypes, derived from 4,115 documents with 60,546 event instances. 
The details of \texttt{OntoEvent} are introduced in appendix. 
We show the main statistics of \texttt{OntoEvent} and compare them with some existing widely-used ED datasets in Table~\ref{tab:dataset}. 

\begin{table}[!htbp]
\centering
\small
\vspace{-2mm}
\resizebox{\linewidth}{!}{
	\begin{tabular}{ c | c c c c c}
	\toprule

	\textbf{Dataset} & \textbf{\#Doc} & \textbf{\#Ins} & \textbf{\#SuperT} & \textbf{\#SubT} & \textbf{\#E-E Rel} \\

	\midrule
	
	ACE 2005 		& 599 & 4,090 & 8 & 33 & None \\
	TAC KBP 2017 	& 167 & 4,839 & 8 & 18  & None \\
	FewEvent 		& - & 70,852 & 19 & 100  & None \\
	MAVEN 			& 4,480 & 111,611 & 21 & 168  & None \\
	$\texttt{OntoEvent}$	& 4,115 & 60,546 & 13 & 100 & 3,804 \\
	
	\bottomrule
	\end{tabular}
}
\vspace{-2mm}
	\caption{Statistics of $\texttt{OntoEvent}$ compared with existing widely-used ED datasets. 
	(Doc: document, Ins: instance, SuperT: supertype, SubT: subtype, E-E Rel: event-event relation.) 
	\label{tab:dataset} }
	
\vspace{-4mm}
\end{table}

\texttt{OntoEvent} is established based on two newly proposed datasets for ED: 
MAVEN \cite{EMNLP2020_MAVEN} and FewEvent \cite{WSDM2020_MetaL-EE_DMBPN}. They are constructed from Wikipedia documents or based on existing event datasets, such as ACE-2005\footnote{\url{http://projects.ldc.upenn.edu/ace/}} and TAC-KBP-2017\footnote{\url{https://tac.nist.gov/2017/KBP/Event/index.html}}. 
In terms of event-event relation annotation in \texttt{OntoEvent}, we jointly use two models: TCR \cite{ACL2018_MATRES} is applied to extract temporal and causal relations, and JCL \cite{EMNLP2020_ERE_H-T} is used for extract hierarchical relations. 
The code of OntoED and \texttt{OntoEvent} dataset can be obtained from Github\footnote{\url{https://github.com/231sm/Reasoning_In_EE}}.

\subsection{Baselines}
For \emph{overall evaluation}, we adopt 
CNN-based model DMCNN \cite{ACL2015_DMCNN}, 
RNN-based model JRNN \cite{NAACL2016_JRNN}, 
and GCN-based model JMEE \cite{EMNLP2018_JMEE}. 
Besides, we adopt BERT-based model AD-DMBERT \cite{EMNLP2019_AD-DMBERT} with adversarial imitation learning. 
We also adopt graph-based models OneIE \cite{ACL2020_OneIE} and PathLM \cite{EMNLP2020_PathLM} which generate graphs from event instances for ED. 
% \noindent 

\noindent For \emph{few-shot evaluation} and \emph{zero-shot evaluation}, we adopt some metric-based models for few-shot ED, such as MatchNet \cite{ACL2020-NUSE_MatchEE}, ProtoNet \cite{NIPS2017_PN} and DMBPN \cite{WSDM2020_MetaL-EE_DMBPN}. 
We also adopt knowledge-enhanced model EKD \cite{ACL2020_EKD} and BERT-based models QAEE \cite{EMNLP2020_QAEE} as well as RCEE \cite{EMNLP2020_RCEE} based on machine reading comprehension. 
Besides, we adopt ZSEE \cite{ACL2018_ZSEE} especially for zero-shot ED.

\begin{table*}[!htbp]
\centering
\small
\resizebox{\linewidth}{!}{
	\begin{tabular}{ c |  c  c  c | c  c  c}
	\toprule

	\multirow{2}*{\textbf{Model}} & \multicolumn{3}{c|}{\textbf{Trigger Identification}} & \multicolumn{3}{c}{\textbf{Event Classification}} \\
	\cline{2-7}
	
	 & $\bm{P}$ & $\bm{R}$ & $\bm{F}$ & $\bm{P}$ & $\bm{R}$ & $\bm{F}$ \\

	\midrule
	
	DMCNN \cite{ACL2015_DMCNN}			
	& 64.65 $\pm$ 0.89 & 64.17 $\pm$ 0.94 & 64.15 $\pm$ 0.91 
	& 62.51 $\pm$ 1.10 & 62.35 $\pm$ 1.12 & 63.72 $\pm$ 0.99 \\

	JRNN \cite{NAACL2016_JRNN}			
	& 65.94 $\pm$ 0.88 & 66.67 $\pm$ 0.95 & 66.30 $\pm$ 0.93 
	& 63.73 $\pm$ 0.98 & 63.54 $\pm$ 1.13 & 66.95 $\pm$ 1.03 \\

	JMEE \cite{EMNLP2018_JMEE}			
	& 70.92 $\pm$ 0.90 & 57.58 $\pm$ 0.96 & 61.87 $\pm$ 0.94 
	& 52.02 $\pm$ 1.14 & 53.80 $\pm$ 1.15 & 68.07 $\pm$ 1.02 \\

	% MOGANED \cite{EMNLP2019_MOGANED}	
	% & - $\pm$ 0. & - $\pm$ 0. & - $\pm$ 0. 
	% & - $\pm$ 0. & - $\pm$ 0. & - $\pm$ 0. \\

	% EE-GCN \cite{EMNLP2020_EE-GCN}		
	% & - $\pm$ 0. & - $\pm$ 0. & - $\pm$ 0. 
	% & - $\pm$ 0. & - $\pm$ 0. & - $\pm$ 0. \\

	AD-DMBERT \cite{EMNLP2019_AD-DMBERT} 
	& 74.94 $\pm$ 0.95 & 72.19 $\pm$ 0.91 & 73.33 $\pm$ 0.97 
	& 67.35 $\pm$ 1.01 & \textbf{73.46 $\pm$ 1.12} & 71.89 $\pm$ 1.03 \\

	OneIE \cite{ACL2020_OneIE} 		
	& 74.33 $\pm$ 0.93 & 71.46 $\pm$ 1.02 & 73.68 $\pm$ 0.97 
	& 71.94 $\pm$ 1.03 & 68.52 $\pm$ 1.05 & 71.77 $\pm$ 1.01 \\

	PathLM \cite{EMNLP2020_PathLM} 
	& 75.82 $\pm$ 0.85 & 72.15 $\pm$ 0.94 & 74.91 $\pm$ 0.92 
	& 73.51 $\pm$ 0.99 & 68.74 $\pm$ 1.03 & 72.83 $\pm$ 1.01 \\

	\textbf{OntoED}	
	& \textbf{77.67 $\pm$ 0.99} & \textbf{75.92 $\pm$ 0.92} & \textbf{77.29 $\pm$ 0.98}
	& \textbf{75.46 $\pm$ 1.06} & 70.38 $\pm$ 1.12 & \textbf{74.92 $\pm$ 1.07} \\	
	
	\bottomrule
	\end{tabular}
}
% \vspace{-1mm}
	\caption{Evaluation of event detection with overall instances. 
	$P(\%)$, $R(\%)$ and $F(\%)$ stand for precision, recall, and F1-score respectively. 
	\label{tab:exp_overall_evaluation} }
	
\end{table*}

\begin{comment}

\begin{table*}[!htbp]
\centering
\small
% \resizebox{\linewidth}{!}{
	\begin{tabular}{ c | c |  c  c  c | c  c  c}
	\toprule

	\multirow{2}*{\textbf{Model}} & \multirow{2}*{\textbf{Encoder}} & \multicolumn{3}{c|}{\textbf{Trigger Identification}} & \multicolumn{3}{c}{\textbf{Event Classification}} \\
	\cline{3-8}
	
	 &  & $\bm{P}$ & $\bm{R}$ & $\bm{F}$ & $\bm{P}$ & $\bm{R}$ & $\bm{F}$ \\

	\midrule
	
	DMCNN \cite{ACL2015_DMCNN}			& CNN	& - & - & - & - & - & - \\
	JRNN \cite{NAACL2016_JRNN}			& RNN	& - & - & - & - & - & - \\
	JMEE \cite{EMNLP2018_JMEE}			& GCN 	& - & - & - & - & - & - \\
	MOGANED \cite{EMNLP2019_MOGANED}	& GCN 	& - & - & - & - & - & - \\
	EE-GCN \cite{EMNLP2020_EE-GCN}		& GCN 	& - & - & - & - & - & - \\
	AD-DMBERT \cite{EMNLP2019_AD-DMBERT}& BERT	& - & - & - & - & - & - \\
	RCEE \cite{EMNLP2020_RCEE}			& BERT	& - & - & - & - & - & - \\
	QAEE \cite{EMNLP2020_QAEE}			& BERT	& - & - & - & - & - & - \\

	\midrule

	OntoED	& CNN	& - & - & - & - & - & - \\
	OntoED	& RNN	& - & - & - & - & - & - \\
	OntoED	& GCN 	& - & - & - & - & - & - \\
	OntoED	& BERT	& - & - & - & - & - & - \\
	
	\bottomrule
	\end{tabular}
% }
% \vspace{-1mm}
	\caption{Overall Evaluation.
	\label{tab:exp_overall_evaluation} }
	
\end{table*}

\end{comment}

\subsection{Experiment Settings}
With regard to settings of the training process, SGD \cite{Ketkar2014_SGD} optimizer is used, with 30,000 iterations of training and 2,000 iterations of testing.
The dimension of token embedding is 50, and the maximum length of a token sequence is 128. In OntoED, a dropout rate of 0.2 is used to avoid over-fitting, and the learning rate is $1 \times 10^{-3}$. 
The hyperparameters of $\gamma$, $\lambda$, $\alpha$, and $\beta$ are set to 0.5, 0.5, 1.5 and 1 respectively. $\psi_S$, $\psi_V$, and $\psi_T$ are set to 0.5, 0.5 and 1 respectively. 
As the dataset is unbalanced, we evaluate the performance of ED with macro precision (P), Recall (R) and adopt micro F1~Score (F) following \cite{ACL2015_DMCNN}. 
Detailed performance can be found in Github\footnotemark[7].
% \footnote{\url{https://github.com/231sm/Reasoning\_In\_EE}}.

% stochastic gradient descent (SGD)

\subsection{Overall Evaluation}
\emph{Setting}. 
We follow the similar evaluation protocol of standard ED models, \eg, DMCNN \cite{ACL2015_DMCNN}. Event instances are split into training, validating, and testing subset with ratio of 0.8, 0.1 and 0.1 respectively. Note that there are no new event types in testing set which are not seen in training. % set

As seen from Table~\ref{tab:exp_overall_evaluation}, OntoED achieves larger gains compared to conventional baselines, \eg, DMCNN, JRNN and JMEE. Moreover, OntoED still generally excel BERT-based AD-DMBERT. This implies the effectiveness of ED framework with ontology embedding, which can leverage and propagate correlations among event types, so that reduce the dependence on data to some extent. 
Especially, OntoED also outperform graph-based models, \ie, OneIE and PathLM. The possible reason is that although they both convert sentences into instance graphs, and PathLM even connects event types with multiple entities, the event correlations are still implicit and hard to capture. OntoED can explicitly utilize event correlations and directly propagate information among event types.

\begin{table*}[!htbp]
\centering
\small
\resizebox{\linewidth}{!}{
	\begin{tabular}{ c | c c c c c}
	\toprule

	\textbf{Model} & \textbf{1\%} & \textbf{5\%} & \textbf{10\%} & \textbf{15\%} & \textbf{20\%} \\

	\midrule

	MatchNet \cite{ACL2020-NUSE_MatchEE}	
	& 7.09 $\pm$ 2.10 & 14.22 $\pm$ 2.18 & 20.59 $\pm$ 2.11 
	& 26.34 $\pm$ 2.07 & 30.93 $\pm$ 1.98 \\ 

	ProtoNet \cite{NIPS2017_PN}				
	& 8.18 $\pm$ 2.51 & 15.25 $\pm$ 2.27 & 21.74 $\pm$ 2.02 
	& 27.75 $\pm$ 2.01 & 32.21 $\pm$ 1.66 \\

	DMBPN \cite{WSDM2020_MetaL-EE_DMBPN}	
	& 11.25 $\pm$ 2.13 & 20.03 $\pm$ 1.99 & 27.69 $\pm$ 1.95 
	& 33.13 $\pm$ 1.91 & 38.06 $\pm$ 1.54 \\

	EKD \cite{ACL2020_EKD}					
	& 35.82 $\pm$ 2.02 & 44.51 $\pm$ 1.83 & 49.64 $\pm$ 1.77 
	& 52.79 $\pm$ 1.26 & 55.95 $\pm$ 1.34 \\

	QAEE \cite{EMNLP2020_QAEE}				
	& 41.17 $\pm$ 1.85 & 48.69 $\pm$ 1.84 & 54.03 $\pm$ 1.81
	& 58.97 $\pm$ 1.81 & 62.04 $\pm$ 1.67 \\

	RCEE \cite{EMNLP2020_RCEE}				
	& 42.58 $\pm$ 1.94 & 50.06 $\pm$ 1.81 & 56.51 $\pm$ 1.71 
	& 60.79 $\pm$ 1.13 & 63.98 $\pm$ 1.08 \\

	\textbf{OntoED}	
	& \textbf{44.98 $\pm$ 1.98} & \textbf{52.19 $\pm$ 1.74} & \textbf{58.53 $\pm$ 1.75}
	& \textbf{62.47 $\pm$ 1.52} & \textbf{65.51 $\pm$ 1.17} \\	
	
	\bottomrule
	\end{tabular}
}
% \vspace{-1mm}
	\caption{F1 score (\%) of event classification on extremely sparse intances for few-shot evaluation. 
	\label{tab:exp_few-shot_evaluation} }
	
\end{table*}

\subsection{Few-shot Evaluation}
\emph{Setting}. 
% To further evaluate the performance of OntoED in few-shot ED tasks, 
We follow the similar evaluation protocol and metrics of data-scarce ED models, \ie, RCEE \cite{EMNLP2020_RCEE}, which train models with partial data. We randomly sample nearly 80\% event types for training, 10\% for validating, and 10\% for testing. Differently from overall evaluation, the event types in testing set are not exsiting in training set.

\begin{figure}[!htbp]
  \centering
  \includegraphics[width=0.98\linewidth]{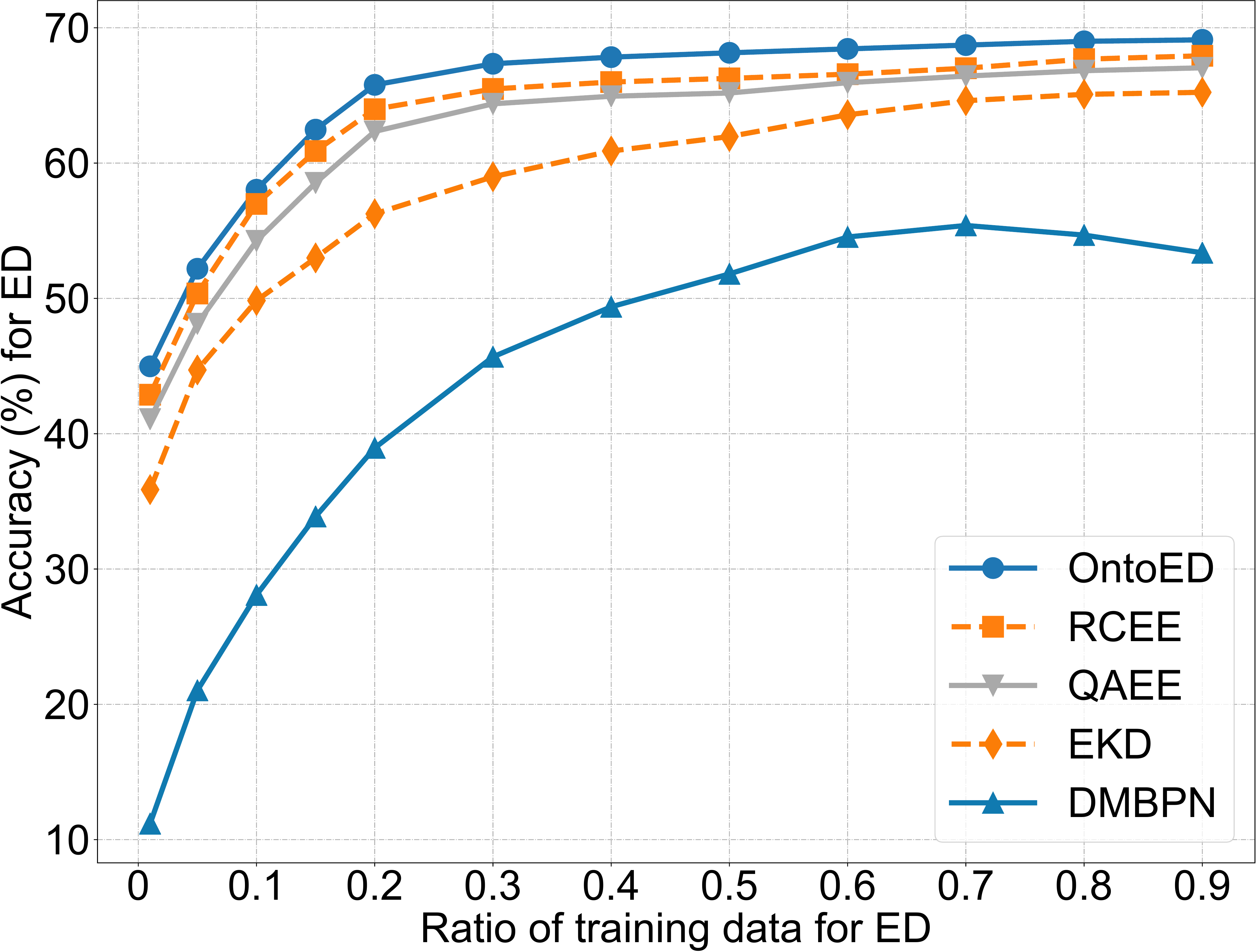}
  \caption{Results on different ratios of ED training data. 
  \label{fig:exp_low_ed} }
  \vspace{-10mm}
\end{figure}

As seen from Table~\ref{tab:exp_few-shot_evaluation}, we demonstrate F1 score results in extremely low-resource scenarios (training with less than 20\% data, with the similar setting to \citet{EMNLP2020_RCEE}). 
Obviously, OntoED behaves tremendous advantages in low-resource ED. 
For example, OntoED obtains 44.98\% F1 with 1\% data, in comparison to 7.09\% in MatchNet and 8.18\% in ProtoNet. 
We also illustrate accuracy results with different ratios of training data followed by \citet{EMNLP2020_RCEE}, show in Figure~\ref{fig:exp_low_ed}. 
As seen, OntoED demonstrates superior performance with less data dependence than baselines. Especially comparing with DMBPN and EKD, which require 60\% training data to closely achieve the best results, while OntoED only uses 20\%. 
Besides, we find that the performance on DMBPN increases first and then slightly decreases as the ratio of training data increases, the possible reason may lie in data noise and redundancy. In low-resource scenarios, more data are not always better. Particularly for some merely data-driven ED models, such as DMBPN, may obtain a worse effect instead if added data are dirty or duplicated. But for OntoED, as it utilizes correlation knowledge in the event ontology and has less dependence on event instances, making it more robust to noisy and redundant data. 
Furthermore, OntoED also outperforms than BERT-based model with regarding each event instance as a question, \ie, QAEE and RCEE. This implies that event ontology learning with event type knowledge may resolve low-resource ED more advantageously than training merely with event instances.

\subsection{Zero-shot Evaluation}
\emph{Setting}. 
% To evaluate the performance of OntoED in zero-shot ED tasks, 
We follow the similar evaluation protocol and metrics of zero-shot ED models, \ie, ZSEE \cite{ACL2018_ZSEE}, and comply with the same dataset segmentation policy as few-shot evaluation, thus there are also new unseen event types for testing. Differently, ED data are completely banned for training, meaning that we train models only with event types other than instances.

\begin{table}[!htbp]
\centering
\small
\resizebox{\linewidth}{!}{
	\begin{tabular}{ c | c c c}
	\toprule

	\textbf{Model} & \textbf{P} & \textbf{R} & \textbf{F} \\

	\midrule
	
	EKD \cite{ACL2020_EKD}		& 32.58 & 31.77 & 32.17 \\
	QAEE \cite{EMNLP2020_QAEE}	& 36.69 & 37.33 & 37.01 \\
	RCEE \cite{EMNLP2020_RCEE}	& 37.45 & 36.83 & 37.14 \\
	ZSEE \cite{ACL2018_ZSEE}	& 40.92 & \textbf{44.18} & 43.02 \\

	\textbf{OntoED}				& \textbf{42.13} & 44.04 & \textbf{43.06} \\
	
	\bottomrule
	\end{tabular}
}
% \vspace{-1mm}
	\caption{Comparisons of performance on zero-shot ED.
	\label{tab:exp_zero-shot_evaluation} }
	\vspace{-8.5mm}
\end{table}

\begin{comment}
\begin{table}[!htbp]
\centering
\small
% \resizebox{\linewidth}{!}{
	\begin{tabular}{ c | c c c c}
	\toprule

	\textbf{Model} & \textbf{Hit@1} & \textbf{Hit@3} & \textbf{Hit@5} & \textbf{Hit@10} \\

	\midrule
	
	ZSEE \cite{ACL2018_TL-EE}	& - & - & - & - \\
	EKD \cite{ACL2020_EKD}		& - & - & - & - \\
	RCEE \cite{EMNLP2020_RCEE}	& - & - & - & - \\
	QAEE \cite{EMNLP2020_QAEE}	& - & - & - & - \\

	OntoED						& - & - & - & - \\
	
	\bottomrule
	\end{tabular}
% }
% \vspace{-1mm}
	\caption{Zero-Shot Evaluation.
	\label{tab:exp_zero-shot_evaluation} }
	
\end{table}\begin{table}[!htbp]
\centering
\small
% \resizebox{\linewidth}{!}{
	\begin{tabular}{ c | c c c c}
	\toprule

	\textbf{Model} & \textbf{Hit@1} & \textbf{Hit@3} & \textbf{Hit@5} & \textbf{Hit@10} \\

	\midrule
	
	ZSEE \cite{ACL2018_TL-EE}	& - & - & - & - \\
	EKD \cite{ACL2020_EKD}		& - & - & - & - \\
	RCEE \cite{EMNLP2020_RCEE}	& - & - & - & - \\
	QAEE \cite{EMNLP2020_QAEE}	& - & - & - & - \\

	OntoED						& - & - & - & - \\
	
	\bottomrule
	\end{tabular}
% }
% \vspace{-1mm}
	\caption{Zero-Shot Evaluation.
	\label{tab:exp_zero-shot_evaluation} }
	
\end{table}
\end{comment}

Table~\ref{tab:exp_zero-shot_evaluation} demonstrates the results regarding zero-shot ED. We can see that OntoED achieves best precision and F1 score as well as comparable recall results in comparison to baselines. 
This illustrates the effectiveness of OntoED handling new unseen event types without introducing outsourcing data. 
% , comparing with baselines. 
Traditional models, such as EKD and RCEE, require to adopt other datasets, \eg, WordNet \cite{J1990_WordNet} (where words are grouped and interlinked with semantic relations) and FrameNet \cite{FSNLP2014_FrameNet} (where frames are treated as meta event types) to increase the persuasiveness of results. 
In contrast, OntoED naturally models the structure of event types with an event ontology, thus even for a new unseen event type without instance data, we can also obtain its representation through the event-event correlation. 
Moreover, OntoED is also beneficial to resolve zero-shot ED than ZSEE. This may due to OntoED modeling with both intra and inter structures of events while ZSEE merely considering the intra-structure.

% #################################
\section{Further Analysis}

\subsection{Ablation Study}
\begin{figure}[!htbp]
  \centering
  \includegraphics[width=0.96\linewidth]{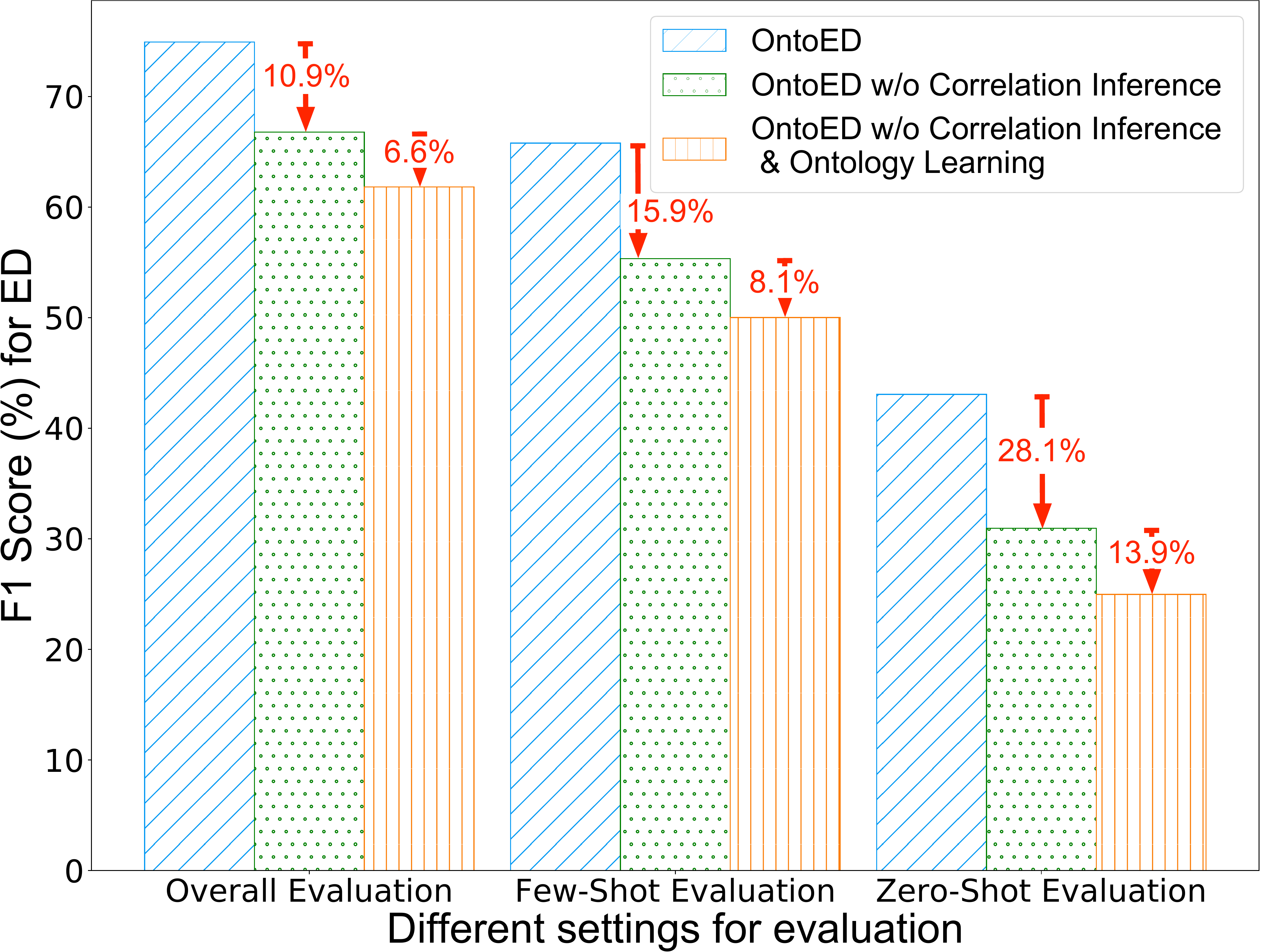}
  \caption{Effect of different modules in OntoED. 
  \label{fig:exp_ed_ablation} }
  \vspace{-4.5mm}
\end{figure}

To assess the effect of event ontology learning and correlation inference, we remove the two modules in OntoED, and evaluate F1 score shown in Figure~\ref{fig:exp_ed_ablation}. 
From the results, we observe that OntoED outperforms the two baselines in all evaluation settings, indicating that event ontology learning and correlation inference facilitate ED, as they utilize knowledge among event types and has less dependence on instance data. 
% , \ie, overall, few-shot and zero-shot
Furthermore, in terms of performance degradation compared to OntoED, F1 score of OntoED merely without event correlation inference (\eg, 10.9\%$\downarrow$) drops more seriously than that without event ontology learning (\eg, 6.6\%$\downarrow$), and the phenomenon is more obvious in few-shot and zero-shot evaluation (\eg, 10.9\%$\downarrow$ v.s. 15.9\%$\downarrow$ and 28.1\%$\downarrow$). 
This illustrates that event correlation inference is more necessary in OntoED, as it establishes more correlations among event types, thereby knowledge can be propagated more adequately, especially from data-rich to data-poor events.

\subsection{Error Analysis}
We further conduct error analysis and provide some representative examples.
(1) One typical error relates to similar event-event structures in the event ontology. As OntoED considers event correlations, event types with similar neighbor triples can be indistinguishable. For example, \emph{Robbery} and \emph{Kidnapping} have the same supertype \emph{Crime}, and they both have the neighbor triples of ($*$, \textsc{Cause}, \emph{Arrest}). 
(2) The second error relates to wrong instance relations. As the instance relation extraction directly influence the establishment of event correlations, wrong instance relations will cause error propagation. 
(3) The third error relates to the same event mention for different event types. For example, ‘\emph{Of the 126 people aboard, 47 died and 74 sustained serious injuries.}’ both mentions \emph{Die} and \emph{Injure}.

\iffalse
\bibliographystyle{acl_natbib}
\bibliography{acl2021}
\fi
% !TEX root = ./acl2021.tex

\section{Conclusion and Future Work}
\label{sec:con_fw}

This paper proposes a novel event detection framework with ontology embedding called \emph{OntoED}. 
We revisit the ED task by linking each event instance to a specific type in a target event ontology. To facilitate the linkage, we enrich the event ontology with event-event relations, such as temporal, causal and hierarchical correlation, and induce more event correlations based on existing ones. 
The key insight is that event ontology can help to reduce model dependence on instance data, especially in low-resource scenarios. As data-rich event types can propagate correlation knowledge to data-poor ones, and new event types can establish linkages to the event ontology. 
We demonstrate the effectiveness of OntoED in three settings: overall, few-shot as well as zero-shot, and experiments show that OntoED excels previous methods with great robustness. 
% particularly with low-resource regimes 

In the future, we intend to extend our work in several aspects. 
First, we would improve the event ontology and consider more event correlations. 
Second, we would explore if low-resource ED can also boost to identify event correlation. 
Third, we would develop more neuro-symbolic methods for ED.

\section*{Acknowledgments}
We want to express gratitude to the anonymous reviewers for their hard work and kind comments. This work is funded by National Key R\&D Program of China (Funding No.2018YFB1402800) and NSFC91846204.

\section*{Broader Impact Statement} 
A broad goal of event detection is to extract structured knowledge from unstructured texts to facilitate knowledge acquisition. For example, it is valuable in the medical domain and provides social benefits to analyze dispensatory details as well as electronic health records. 
Furthermore, a solid ED system can also be applied to many society issues, such as anti-terrorist and public opinion analysis. 

In this paper, we present a new dataset \texttt{OntoEvent} for ED with event-event correlations. 
The event data are all collected from existing datasets (\ie, ACE 2005) or open source databases (\eg, Wikipedia), and the annotation are generated from existing models with citations. 
In experiments, we detailedly describe how to evaluate the newly-proposed \texttt{OntoEvent} and provide specific analysis. 
The code and dataset are both available. 

Our approach to ED can leverage only a few event corpus to establish the linkage between event types and event instances \wrt event correlations. 
In addition, this work is also a brand-new attempt to combine information extraction and symbolic reasoning, based on ontology embedding. 
Our intention is to develop an ontology-based ED system for the NLP community, and wish our innovation can become a small step in this direction.

% \clearpage

\balance
\bibliographystyle{acl_natbib}
\bibliography{acl2021}

\clearpage

\appendix

% \section{Hierarchical Event Type Schema}
\section{Hierarchical Event Type Schema ~~ (Part 1)}

% Here we list detailed statistics of \texttt{OntoEvent}.

\begin{table}[!htbp]
\centering
% \footnotesize
\scriptsize
% \resizebox{\linewidth}{!}{
	\begin{tabular}{c | c | c}
	\toprule
	  
	\textbf{SuperType} & \textbf{SubType} & \textbf{\#Instance} \\
	
	\midrule

	\multirow{5}*{Business}
	 & Collaboration & $272$ \\ 
	 & Declare-Bankruptcy & $95$ \\ 
	 & End-Org & $115$ \\ 
	 & Merge-Org & $41$ \\ 
	 & Start-Org & $162$ \\ 

	\midrule

	\multirow{7}*{Cause-Effect}
	 & Causation & $3387$ \\ 
	 & Cause-Change-Of-Position-On-A-Scale & $1138$ \\ 
	 & Cause-Change-Of-Strength & $1180$ \\ 
	 & Cause-To-Amalgamate & $374$ \\ 
	 & Cause-To-Be-Included & $850$ \\ 
	 & Cause-To-Make-Progress & $448$ \\ 
	 & Influence & $812$ \\ 

	\midrule

	\multirow{5}*{Commerce}
	 & Carry-Goods & $62$ \\ 
	 & Commerce-Buy & $64$ \\ 
	 & Commerce-Pay & $114$ \\ 
	 & Commerce-Sell & $153$ \\ 
	 & Manufacturing & $411$ \\ 

	\midrule

	\multirow{18}*{Conflict}
	 & Attack & $3550$ \\ 
	 & Bearing-Arms & $115$ \\ 
	 & Besieging & $312$ \\ 
	 & Conquering & $1754$ \\ 
	 & Defending & $923$ \\ 
	 & Escaping & $956$ \\ 
	 & Hostile-Encounter & $3620$ \\ 
	 & Killing & $2030$ \\ 
	 & Military-Operation & $1257$ \\ 
	 & Protest & $185$ \\ 
	 & Quarreling & $215$ \\ 
	 & Releasing & $143$ \\ 
	 & Rescuing & $162$ \\ 
	 & Revenge & $64$ \\ 
	 & Sending & $512$ \\ 
	 & Terrorism & $309$ \\ 
	 & Use-Firearm & $538$ \\ 
	 & Violence & $419$ \\ 

	\midrule
	
	\multirow{6}*{Contact}
	 & Broadcast & $638$ \\ 
	 & Come-Together & $400$ \\ 
	 & Communication & $582$ \\ 
	 & Contact & $237$ \\ 
	 & Correspondence & $197$ \\ 
	 & Telling & $195$ \\ 

	\midrule

	\multirow{3}*{Crime}
	 & Kidnapping & $109$ \\ 
	 & Robbery & $70$ \\ 
	 & Theft & $24$ \\

	\midrule

	% \multirow{16}*{Justice}
	\multirow{16}*{Justice}
	 & Acquit & $27$ \\ 
	 & Appeal & $125$ \\ 
	 & Arrest & $379$ \\ 
	 & Committing-Crime & $265$ \\ 
	 & Convict & $260$ \\ 
	 & Criminal-Investigation & $277$ \\ 
	 & Execute & $70$ \\ 
	 & Extradition & $14$ \\ 
	 & Fine & $84$ \\ 
	 & Justifying & $34$ \\ 
	 & Legal-Rulings & $241$ \\ 
	 & Pardon & $10$ \\ 
	 & Prison & $69$ \\ 
	 & Release-Parole & $95$ \\ 
	 & Sentence & $321$ \\ 
	 & Sue & $196$ \\ 

	\midrule

	\multirow{11}*{Life}
	% \multirow{6}*{Life}
	 & Award & $71$ \\ 
	 & Be-Born & $128$ \\ 
	 & Bodily-Harm & $1255$ \\ 
	 & Breathing & $7$ \\ 
	 & Cure & $88$ \\ 
	 & Death & $1226$ \\ 
	 & Divorce & $65$ \\ 
	 & Education-Teaching & $132$ \\ 
	 & Marry & $205$ \\ 
	 & Name-Conferral & $760$ \\ 
	 & Recovering & $337$ \\ 

	\bottomrule
	\end{tabular}
% }
% \vspace{-3mm}
\caption{Hierarchical event types in \texttt{OntoEvent}. 
\label{tab:Hier_Event_Types_1}}
% \vspace{-3mm}
\end{table}

\section{Hierarchical Event Type Schema ~~ (Part 2)}

\begin{table}[!htbp]
\centering
% \footnotesize
\scriptsize
% \resizebox{\linewidth}{!}{
	\begin{tabular}{c | c | c}
	\toprule

	\textbf{SuperType} & \textbf{SubType} & \textbf{\#Instance} \\

	\midrule 

	\multirow{8}*{Movement}
	 & Arriving & $1542$ \\ 
	 & Body-Movement & $140$ \\ 
	 & Departing & $488$ \\ 
	 & Motion-Directional & $869$ \\ 
	 & Placing & $815$ \\ 
	 & Transport-Artifact & $241$ \\ 
	 & Transport-Person & $446$ \\ 
	 & Traveling & $937$ \\ 

	\midrule

	\multirow{3}*{Natural-Disaster}
	 & Catastrophe & $3974$ \\ 
	 & Damaging & $1227$ \\ 
	 & Destroying & $1445$ \\ 

	\midrule

	\multirow{6}*{Personnel}
	 & Change-Of-Leadership & $933$ \\ 
	 & Elect & $577$ \\ 
	 & Employment & $156$ \\ 
	 & End-Position & $887$ \\ 
	 & Nominate & $45$ \\ 
	 & Start-Position & $506$ \\ 

	\midrule

	\multirow{4}*{Proecess}
	 & Confronting-Problem & $225$ \\ 
	 & Process-End & $1730$ \\ 
	 & Process-Start & $3275$ \\ 
	 & Resolve-Problem & $157$ \\ 

	\midrule

	\multirow{8}*{Transaction}
	 & Earnings-And-Losses & $1104$ \\ 
	 & Exchange & $79$ \\ 
	 & Getting & $1067$ \\ 
	 & Giving & $695$ \\ 
	 & Receiving & $362$ \\ 
	 & Renting & $21$ \\ 
	 & Supply & $618$ \\ 
	 & Transaction & $50$ \\ 

	\bottomrule
	\end{tabular}
% }
% \vspace{-3mm}
\caption{Hierarchical event types in \texttt{OntoEvent}. 
\label{tab:Hier_Event_Types_2}}
% \vspace{-3mm}
\end{table}

\clearpage

\section{Overview of Temporal \& Causal Event-Event Correlations}

\begin{table}[!htbp]
\centering
% \scriptsize
\footnotesize
% \resizebox{\linewidth}{!}{
	\begin{tabular}{c | c | c }
	\toprule
	  
	\textbf{E-E Relation} & \textbf{Head Event Type} & \textbf{Tail Event Type} \\
	% & \textbf{\#Instance}

	\midrule

	\multirow{18}*{\textsc{Before}}
	 & Personnel.Start-Position  & Personnel.End-Position  \\
	 & Personnel.Nominate  & Personnel.Elect  \\
	 & Commerce.Commerce-Sell  & Commerce.Commerce-Buy  \\
	 & Commerce.Manufacturing  & Commerce.Carry-Goods  \\
	 & Life.Marry  & Life.Divorce  \\
	 & Life.Be-Born  & Life.Name-Conferral  \\
	 & Justice.Arrest  & Justice.Prison  \\
	 & Justice.Sue  & Justice.Criminal-Investigation  \\
	 & Justice.Criminal-Investigation  & Justice.Legal-Rulings \\
	 & Transaction.Transaction  & Transaction.Earnings-And-Losses  \\
	 & Proecess.Confronting-Problem  & Proecess.Resolve-Problem  \\
	 & Justice.Justifying  & Justice.Committing-Crime  \\
	 & Justice.Convict  & Justice.Execute  \\
	 & Justice.Convict  & Justice.Fine  \\
	 & Justice.Convict  & Justice.Extradition  \\
	 & Justice.Convict  & Justice.Sentence  \\
	 & Justice.Sentence  & Justice.Release-Parole  \\
	 & Justice.Acquit  & Justice.Pardon  \\

	\midrule

	\multirow{10}*{\textsc{After}}
	 & Movement.Arriving & Movement.Transport-Artifact  \\
	 & Movement.Arriving & Movement.Departing  \\
	 & Movement.Arriving & Movement.Transport-Person \\
	 & Business.End-Org  & Business.Start-Org  \\
	 & Life.Death  & Life.Be-Born  \\
	 & Life.Cure  & Life.Bodily-Harm  \\
	 & Proecess.Process-End  & Proecess.Process-Start  \\
	 & Justice.Appeal  & Justice.Sue  \\
	 & Conflict.Escaping  & Conflict.Besieging  \\
	 & Conflict.Bearing-Arms  & Conflict.Use-Firearm  \\

	\midrule

	\multirow{20}*{\textsc{Equal}}
	 & Commerce.Commerce-Pay  & Commerce.Commerce-Buy  \\
	 & Business.Declare-Bankruptcy  & Business.End-Org  \\
	 & Business.Merge-Org  & Business.Collaboration  \\
	 & Transaction.Getting  & Transaction.Receiving  \\
	 & Transaction.Giving  & Transaction.Supply  \\
	 & Transaction.Renting  & Transaction.Exchange  \\
	 & Movement.Traveling  & Movement.Transport-Person \\
	 & Natural-Disaster.Damaging  & Natural-Disaster.Destroying  \\
	 & Movement.Body-Movement  & Movement.Traveling  \\
	 & Life.Cure  & Life.Recovering  \\
	 & Justice.Extradition  & Justice.Legal-Rulings  \\
	 & Conflict.Revenge  & Conflict.Hostile-Encounter  \\
	 & Conflict.Protest  & Conflict.Quarreling  \\
	 & Conflict.Protest  & Conflict.Use-Firearm  \\
	 & Conflict.Protest  & Conflict.Violence  \\
	 & Conflict.Protest  & Conflict.Attack  \\
	 & Conflict.Protest  & Conflict.Killing  \\
	 & Conflict.Protest  & Conflict.Besieging  \\
	 & Conflict.Protest  & Conflict.Conquering  \\
	 & Conflict.Protest  & Conflict.Defending  \\

	\midrule

	\multirow{4}*{\textsc{Cause}}
	 & 	Cause-Effect.Causation & Cause-Effect.Influence \\
	 & 	Natural-Disaster.Catastrophe & Natural-Disaster.Damaging \\
	 & 	Conflict.Attack & Life.Bodily-Harm \\
	 & 	Conflict.Killing & Life.Death \\

	\midrule

	\multirow{5}*{\textsc{CausedBy}}
	 & Justice.Arrest  & Crime.Kidnapping  \\
	 & Justice.Arrest  & Crime.Robbery  \\
	 & Justice.Arrest  & Crime.Theft  \\
	 & Justice.Arrest  & Conflict.Attack  \\
	 & Justice.Arrest  & Conflict.Killing  \\

	\bottomrule
	\end{tabular}
% }
% \vspace{-3mm}
\caption{Overview of temporal \& causal event-event correlations in \texttt{OntoEvent}. 
\label{tab:Event_Event_Rels}}
% \vspace{-3mm}
\end{table}

\end{document}